# Various Decays of Particles, Universal Decay Formulas and Their Possible Dynamic Basis and Applications

Yi-Fang Chang
Department of Physics, Yunnan University, Kunming 650091, China
(E-mail: yifangchang1030@hotmail.com)

**Abstract:** The decay modes and fractions in particle physics are some quantitative and very complex questions. Various decays of particles and some known decay formulas are discussed. Many important decays of particles ($0^{-+}$ mesons, $(1/2)^{+}$ baryons and $\mu^{\pm}$) and some known decays of resonances can be generally described by the square of some types of the associated Legendre functions $P_l^l$ and $P_{l+1}^l$, $\Gamma = \frac{G^2}{4\pi N} m_i^{\alpha} |P_l^{|m|}(x)|^2$, in which $x = \frac{\sum m_f}{m_i}$. It is combined with the $\Delta$I=1/2 rule, then the agreements are very good. The decay formulas of the similar decay modes are the same. For the same types of particles, the coupling constants are kept to be invariant, and only six constants are independent. The simplest dynamical basis is probably that the decay formulas are the square of the solution of free field equation of the decay particles in the momentum coordinate system. Further, we discuss some general decays and their rules. Finally, we apply the decay formulas to some massive hadrons.

## 1. Introduction

A very difficult question in particle physics is the diversity and the complexity on the decay modes and the fractions of particles. For example, there are twenty-two quantitative modes and the total forty-nine modes on $K^{\pm}$, and difference of these fractions amounts to $10^{11}$ times [1]. Theoretically all of decay rate can be obtained, and usual decay theory is to apply the quantum field theory. Some models combine the conservation laws, the selection rules and symmetry. According to various decays, the theory may derive the corresponding current-current interactions, and their current algebra and Lagrangian [2,3]. The theory agrees better for the leptonic and semileptonic decays. A well-known theory is the Cabibbo theory [4].

Feynman and Gell-Mann discussed theory of the Fermi interaction, including $\beta$ decay and the weak decays of strange particles and so on [5]. Cabibbo investigated the unitary symmetry and leptonic decays, and derived the Cabibbo angle [4]. Chounet, et al., reviewed the leptonic decays of hadrons [6]. Smith and Uy discussed the low-energy theorem for $\gamma \to 3\pi$, and derived a unitarity bound on the decay rate for $K \to \pi\gamma\gamma$ [7]. Holstein studied the recoil effects in allowed beta decay [8]. Bramon and Greco discussed the radiative decays of mesons in the quark model [9]. Bajaj, et al., presented a gauge-invariant and renormalizable model for the nonleptonic decays of mesons and baryons, and calculated the two-pion and three-pion decays of the kaon and the s-wave amplitudes for the hyperon decays. These results for $K \to 2\pi$ and $K \to 3\pi$ decay widths are in excellent agreement with experiment [10]. Abe, et al., investigated the vector-like model and nonleptonic decays of ordinary and charmed hadrons [11]. Bryman and Picciotto reviewed the double beta decay [12]. Sirlin developed a current algebra formulation of the radiative corrections in gauge theories and the universality of the weak interactions, and given the analogous calculation for the transition rate of pion decay [13]. Greco and Srivastava discussed some the decay formulas [14]. Ozaki discussed decay of $K_{l4}$ in a relativistic quark model [15]. Devlin and Dickey reviewed the weak hadronic decays $K \to 2\pi$ and $K \to 3\pi$ with the CP conserving transitions [16]. Donnell reviewed the radiative decays of mesons and some developments [17]. Minami and Nakashima investigated the decays of pseudoscalar mesons [18,19]. Kaxiras, et al., calculated the radiative decay widths of the low-lying strange baryons on both within the relativistic quark bag model and the nonrelativistic potential model [20].

The concrete models and the last formulas, which are derived through a host of calculation, are manifold. For the $M \to l\nu$ and $K \to 2\pi$, the decay-formulas of the field theory are able to be developed. But, many results disagree with the experimental values, or may usually be owing to different coupling constants. The numbers of the authors and the references are all immense, but the majority is only discussed a type of decays, even a mode of a particle. Some became more and more complex, and many parameters are introduced. Some are incongruity with the experiments. Further, the method of the S-matrix, which derives the decay rate, always is approximate. Of course, many theories of particle physics do not consider and calculate these very complex questions.

We proposed the unified decay formulas for hadrons and $\mu^{\pm}$ [21]. Then we researched systematically the rules of decay modes and fractions for the more stable particles, and obtained four qualitative relations among decay modes, and expounded quantitatively the known decay probability formulas and their extension, some decay widths with ratio and in proportion to power of the mass-difference, the applications of $\Delta$I=1/2 rule, the usual fractions with ratio of equality for the particles of many decay modes and so on [22].

Ritchie and Wojcicki studied some rare K decays [23]. Richman and Burchat reviewed the experimental measurements and theoretical descriptions of leptonic and semileptonic decays of charm and bottom hadrons [24]. Surguladze and Samuel investigated the decay widths and total cross sections in perturbative Quantum Chromodynamics (QCD) [25]. Buchalla, et al., reviewed the status of QCD corrections to weak decays beyond the leading-logarithmic approximation, including particle-antiparticle mixing and rare and CP-violating decays, and discuss in detail the effective Hamiltonians of all decays [26]. Kuno and Okada reviewed the current theoretical and experimental status of the field of muon decay and its potential to search for new physics beyond the standard model [27]. Hurth discussed present status of inclusive rare B decays [28]. Severijns, et al., reviewed precision tests of the standard electroweak model in nuclear beta decay [29]. Davier, et al., reviewed the experimental results and theoretical predictions of hadronic tau decays [30].

Recently, Ibrahim and Nath discussed CP violation from the standard model and supersymmetry to extra dimensions and string theory, which are related with K and B decays, electric dipole moments, neutrino physics, the baryon asymmetry, collider physics and so on [31]. CLEO Collaboration searched the decay $J/\psi \to \gamma$ +invisible [32]. The Belle Collaboration searched the leptonic decays of $D^0$ mesons [33]. Bozzi, et al., considered the next-to-leading order QCD corrections to $W^+W^-\gamma$ and $ZZ\gamma$ production with leptonic decays [34]. Leitner, et al., discussed K* resonance effects on direct CP violation in $B \to \pi\pi K$ [35]. Davidson and Grenier studied lepton flavor violating Higgs bosons and the rare decay $\tau \to \mu\gamma$ [36]. BABAR Collaboration searched the neutrinoless, lepton-flavor violating decays of the τ lepton into three charged leptons [37]. Briere, et al., analyzed $D^+ \to K^-\pi^+e^+\nu_e$ and $D^+ \to K^-\pi^+\mu^+\nu_\mu$ semileptonic decays [38]. Rodejohann revisited the process of inverse neutrinoless double beta decay, including neutrinos, Higgs triplets and a muon collider [39]. Cheng and Chua reanalyzed the $B \to M$ tensor form factors in a covariant light-front quark model, where M represents a vector meson V, an axial-vector meson A, or a tensor meson T. They discussed radiative B decays [40]. Dubnicka, et al., applied a relativistic constituent quark model with infrared confinement to describe the tetraquark state X(3872) meson, and calculated the decay widths of some observed channels [41].

## 2. Some known decay formulas

So far, the lifetimes of hadrons cannot usually be derived from a unified formula. Only some formulas are obtained for few particles, for example, the muon lifetime $\tau_\mu$ [1-3,42-44].

For $\mu^\pm \to e^\pm \nu_e \nu_\mu$, a formula is [42]:

$$\tau_\mu^{-1} = \frac{G^2 m_\mu^5}{192\pi^3}\left(1-\frac{8m_e^2}{m_\mu^2}\right). \quad (1)$$

Another formula is [3]:

$$\tau_\mu^{-1} = \frac{G^2 m_\mu^5}{192\pi^3}\left[1-\frac{\alpha}{2\pi}\left(\pi^2-\frac{25}{4}\right)\right]. \quad (2)$$

Another exact formula is [44]:

$$\frac{1}{\tau_\mu} = W_\mu = \frac{G^2 m_\mu^5}{192\pi^3}[1-8y+8y^3-y^4-12y^2\ln y]. \quad (3)$$

Here $y \equiv m_e^2/m_\mu^2$ is very small, and $G = (1.16637 \pm 0.00002)\times 10^{-11} MeV^{-2}$ is obtained. We proposed that the unified formula for this decay is [21]:

$$\tau_\mu^{-1} = \frac{G^2 m_\mu^5}{192\pi^3}\left(1-\frac{m_e^2}{m_\mu^2}\right)^6. \quad (4)$$

The three formulas (1)-(4) are different. But, when an electron-mass is neglected, they become a usual formula [1,43].

A semileptonic decay formula is [3,43-45]:

$$\tau_{\pi l2}^{-1}=\Gamma(\pi^{\pm}\to l^{\pm}\nu)=\frac{g^2}{4\pi}m_l^2 m_\pi(1-\frac{m_l^2}{m_\pi^2})^2 . \qquad (5)$$

The formula (1) is similar with the formula of a decay $K^{\pm}\to\pi^0 e^{\pm}\nu_e$ [43]:

$$\Gamma=C\frac{G^2 m_{K^\pm}^2}{\pi^3}(1-\frac{8m_{\pi^0}^2}{m_{K^\pm}^2}) . \qquad (6)$$

Moreover, some known lifetime formulas are:

$$\tau_n^{-1}=2(1+3\alpha^2)\frac{G^2}{\pi^3}m_n^5(1-x^2)^6 = 903.826\text{s; (here } \alpha = 1.21). \qquad (7)$$

$$\tau^{-1}(\pi^{\pm})=\frac{g^2}{4\pi}m(\pi)x^2(1-x^2)^2 . \qquad (8)$$

Here $$x=\frac{\sum m_f}{m_i} , \qquad (9)$$

in which $m_i$ is mass of a particle of the initial state and $\sum m_f$ is the mass-sum of the particles of the final state. For $\tau^{-1}(\pi^0)$,

$$\tau(\pi^{\pm})/\tau(\pi^0)=(2.60/0.84)\times 10^8 = 3\times 10^8 \approx \alpha^{-4} . \qquad (10)$$

$$\tau(K_L^0)/\tau(K_S^0)=(5.114/0.90)\times 10^2 = 5.68\times 10^2 . \qquad (11)$$

For $\tau^{-1}(\Sigma^0)$,

$$\tau(\Sigma^+)/\tau(\Sigma^0)=(8.018/7.4)\times 10^9 = 1.084\times 10^9 . \qquad (12)$$

$$\tau^{-1}(\Sigma^-)=\frac{g_h^2}{4\pi}m(\Sigma)(1-x^2) . \qquad (13)$$

$$\tau^{-1}(\Xi^0)=\frac{1.16 g_h^2}{4\pi}m(\Xi^0)(1-x^2) ; \quad \tau^{-1}(\Xi^-)=\frac{g_h^2}{2\pi}m(\Xi^-)(1-x^2) . \qquad (14)$$

Probably, the formulas of various decays, at least $A\to Be^-\overline{\nu}_e$, should be modified. The width formulas of the two body radiative decay are [9]:

$$\Gamma(V\to P\gamma)=\frac{g_{VP\gamma}^2}{96\pi}m_V^3(1-\frac{m_P^2}{m_V^2})^3 , \qquad (15)$$

$$\Gamma(P \to V\gamma) = \frac{g_{PV\gamma}^2}{32\pi} m_P^4 (1 - \frac{m_V^2}{m_P^2})^3 . \qquad (16)$$

The decay formulas of the two photon are [14]:

$$\Gamma(S \to \gamma\gamma) = \frac{e^4 g_S^2}{16\pi} m_S^3 , \qquad (17)$$

$$\Gamma(P \to \gamma\gamma) = \frac{e^4 g_P^2}{64\pi} m_P^3 . \qquad (18)$$

Oakes and Nandy, et al., discussed $\eta \to 3\pi^0$ decay [46,47]:

$$\Gamma(\eta \to 3\pi^0) = \frac{\sqrt{3}}{9216\pi^2} (\frac{\sin\theta}{f_\pi})^4 m_\eta^5 (1 - \frac{3m_\pi}{m_\eta})^2 . \qquad (19)$$

The leptonic decay formulas are:

$$\Gamma(\varphi \to l\bar{\nu}) = \frac{f^2 [m]^6}{4\pi M^3} , \text{ and } \Gamma(V \to l\bar{l}) = \frac{e^2 \xi^2}{12\pi M^3} . \qquad (20)$$

In quantum field theory the strong decay formula is:

$$\Gamma(N^* \to N\pi^0) = \frac{g^2 [m]^4}{16\pi M^{*3}} . \qquad (21)$$

Eqs.(20)(21) are the two-body decays, and are inverse ratio with cube of mass M. Their forms are similar with an excited radiation coefficient

$$B_{nm} = \frac{c^3 \pi^2}{\hbar\omega^3} A_{nm} . \qquad (22)$$

Eq.(22) is based on the equilibrium conditions of thermodynamics proposed by Einstein.

Theoretically, the modification is smaller or bigger for $e^+e^- / \mu^+\mu^-$. For instance [3],

$$\Gamma(V^0 \to l^+ l^-) \approx \frac{4\pi}{3} e^2 f_{V\gamma}^2 m_V (1 + \frac{2m_i^2}{m_V^2}) . \qquad (23)$$

But, the observed results [1] are the same for $\rho(770), \phi(1020) \to e^+e^- / \mu^+\mu^-$, etc.

Moreover, Jacob and Sachs discussed mass and lifetime of unstable particles [48]. Bander, et al., proposed a mechanism, in which the reaction $D^0 \to$s+d+gluon as a source for the difference in the lifetimes of the charged and neutral D mesons [49]. Sehgal and Leusen investigated the time evolution of decay spectrum in $K^0, \overline{K}^0 \to \pi^+\pi^- e^+ e^-$ [50]. Cabibbo, et al., discussed various semileptonic hyperon decays and Cabibbo-Kbayashi-Maskawa unitarity [51].

## 3. Universal decay formulas

We proposed many important decays of particles ($0^{-+}$ mesons, $(1/2)^{+}$ baryons and $\mu^{\pm}$) and some known decays of resonances can be generally described by the square of some types of the associated Legendre functions $P_l^l$ and $P_{l+1}^l$ [21], and the unified decay formulas are:

$$\Gamma = \frac{G^2}{4\pi N} m_i^{\alpha} \, | P_l^{|m|}(x) |^2 . \qquad (24)$$

Here a variable x is the same with Eq.(9). It combines the selection rules of decays $|\Delta S|$=1 and $|\Delta I| = |\Delta I_3|$=1/2, then the agreements are very good. The decay formulas of the similar decay modes are the same. For the same types of particles, the coupling constants are kept to be invariant, and only six constants are independent [21].

The unified formulas (24) determine completely the fractions of different modes of a particle. The two-body decays of hyperons and mesons are $|P_1^1|^2$ and $|P_3^2|^2$, respectively, and both three-body decays are all $|P_2^2|^2$, $A' \to Ae^{-}\bar{\nu}_e(\Delta S = 0)$ are all $|P_6^6|^2$ and so on. For similar decay modes, the decay formulas is mainly the same, where N is constant of the corresponding $\Delta$I=1/2 rule, and usual N=1. It is consistent with some known results. The coupling constants of the same types of particles are kept to be invariant.

For $H \to Bl\nu$, $(g'_h)^2 = 1.64\times10^{-16}\times(m_H/m_p)^3$. For $\pi^{\pm}$ meson, $g_0^2 = 2.18\times10^{-14}$; for K, $g_1^2 = 2.06\times10^{-14}$. For $A' \to Ae^{-}\bar{\nu}_e\ (\Delta S = 0)$, $G_0 = 1.02\times10^{-5}\times m_p^{-2}$. Here $g_0, g_1$ and $G_0$ are known coupling constants.

For $|\Delta S|$=1, a main decay is $A' \to A\pi$, its width formula is:

$$\Gamma = C|\psi_{1,1}|^2 = \frac{G^2}{4\pi N} m_i (1 - x^2) . \qquad (25)$$

Eq.(25) includes some two-body decays of hyperons $H \to B\pi$: $\Lambda \to p\pi^{-}, \to n\pi^{0}$ (for after decay N=2); $\Sigma^{+} \to p\pi^{0}, \to n\pi^{+}$; $\Sigma^{-} \to n\pi^{-}$; $\Xi^{0} \to \Lambda\pi^{0}$ and $\Xi^{-} \to \Lambda\pi^{-}$ (here N=1/2). Here $G = g_h = 6.2\times10^{-7}\times(m_p/m_H)$. It also includes the radiative decays of K: $K^{\pm} \to \pi^{0}\pi^{\pm}\gamma, \to \pi^{\pm}\pi^{+}\pi^{-}\gamma$, here $G = g_1\alpha/\sqrt{2}$.

| | Calculated | Observed(MeV) |
|---|---|---|
| $\Lambda \to p\pi^-$ | $1.606\times10^{-12}$ | $(1.599\pm0.013)\times10^{-12}$ |
| $\Lambda \to n\pi^0$ | $0.873\times10^{-12}$ | $(0.896\pm0.013)\times10^{-12}$ |
| $\Sigma^+ \to p\pi^0$ | $4.205\times10^{-12}$ | $(4.233\pm0.025)\times10^{-12}$ |
| $\Sigma^+ \to n\pi^+$ | $4.003\times10^{-12}$ | $(3.966\pm0.025)\times10^{-12}$ |
| $\Sigma^- \to n\pi^-$ | $4.222\times10^{-12}$ | $(4.444\pm0.000)\times10^{-12}$ |
| $\Xi^0 \to \Lambda\pi^0$ | $1.955\times10^{-12}$ | $(2.259\pm0.000)\times10^{-12}$ |
| $\Xi^- \to \Lambda\pi^-$ | $3.979\times10^{-12}$ | $(4.011\pm0.000)\times10^{-12}$ |
| $K^\pm \to \pi^0\pi^\pm\gamma$ | $1.488\times10^{-17}$ | $(1.462\pm0.085)\times10^{-17}$ |
| $K^\pm \to \pi^\pm\pi^+\pi^-\gamma$ | $0.605\times10^{-17}$ | $(0.553\pm0.213)\times10^{-17}$ |

For the two-body decays of mesons, their width formula is:

$$\Gamma = C\,|\psi_{3,2}|^2 = \frac{G^2}{4\pi N} m_i x^2 (1-x^2)^2 . \tag{26}$$

Eq.(26) includes $\pi^\pm \to l\nu$, here $G = g_0$; $K^\pm \to l\nu$, here $G = g_1$;

$K_S^0 \to \pi^+\pi^-, \to \pi^0\pi^0$ (N=2), here $G = 2g_1\alpha\times443$; $K^\pm \to \pi^\pm\pi^0$, here $G = 2g_1\alpha\sqrt{443}$;

and $K_L^0 \to \pi^+\pi^-, \to \pi^0\pi^0$ (N=2), here $G = 2g_1\alpha$ .

| | Calculated | Observed(MeV) |
|---|---|---|
| $\pi^\pm \to \mu\nu$ | $2.528\times10^{-14}$ | $(2.528\pm0.002)\times10^{-14}$ |
| $\pi^\pm \to e\nu$ | $3.246\times10^{-18}$ | $(3.110\pm0.058)\times10^{-18}$ |
| $K^\pm \to \mu\nu$ | $3.375\times10^{-14}$ | $(3.372\pm0.009)\times10^{-14}$ |
| $K^\pm \to e\nu$ | $8.671\times10^{-19}$ | $(8.238\pm0.479)\times10^{-19}$ |
| $K_S^0 \to \pi^+\pi^-$ | $5.041\times10^{-12}$ | $(5.084\pm0.018)\times10^{-12}$ |
| $K_S^0 \to \pi^0\pi^0$ | $2.500\times10^{-12}$ | $(2.255\pm0.018)\times10^{-12}$ |
| $K^\pm \to \pi^\pm\pi^0$ | $1.127\times10^{-14}$ | $(1.112\pm0.008)\times10^{-14}$ |
| $K_L^0 \to \pi^+\pi^-$ | $2.568\times10^{-17}$ | $(2.543\pm0.063)\times10^{-17}$ |
| $K_L^0 \to \pi^0\pi^0$ [7] | $1.273\times10^{-17}$ | $(1.119\pm0.229)\times10^{-17}$ |

For the three-body decays of mesons and hyperons, their width formulas are:

$$\Gamma = C|\psi_{2,2}|^2 = \frac{G^2}{4\pi N} m_i (1-x^2)^2 . \quad (27)$$

Eq.(27) includes $K^\pm \to \pi^0 l\nu$ (N=2), $K_L^0 \to \pi^\pm l\nu$ (N=1), here $G = g_1\sqrt{\alpha}$ ;

$K^\pm \to \pi^\pm\pi^+\pi^-, \to \pi^0\pi^0\pi^\pm (N=4)$, here $G = 2.53 g_1\sqrt{\alpha}$ ;

$K_L^0 \to 3\pi^0, \to \pi^+\pi^-\pi^0 (N=3/2)$, here $G = 2 g_1\sqrt{\alpha}$ ;

$K^\pm \to \pi^\pm\pi^\mp l^\pm \nu, \to \pi^0\pi^0 l^\pm \nu (N=2)$ and $K_L^0 \to \pi^0\pi^\pm l^\mp \nu$, here $G = g_1\alpha/\pi$ .

Eq.(27) also includes $H \to Blv(\Delta S = 1)$, i.e., $\Lambda \to plv$ and $\Sigma^- \to nlv$, here $G = g_h'$ .

| | Calculated | Observed(MeV) |
|---|---|---|
| $K^\pm \to \pi^0 e\nu$ [43] | $2.525\times10^{-15}$ | $(2.565\pm0.027)\times10^{-15}$ |
| $K^\pm \to \pi^0 \mu\nu$ | $1.716\times10^{-15}$ | $(1.764\pm0.048)\times10^{-15}$ |
| $K_L^0 \to \pi^\pm e\nu$ | $5.047\times10^{-15}$ | $(5.217\pm0.063)\times10^{-15}$ |
| $K_L^0 \to \pi^\pm \mu\nu$ | $3.413\times10^{-15}$ | $(3.478\pm0.032)\times10^{-15}$ |
| $K^\pm \to \pi^\pm\pi^+\pi^-$ [10] | $2.977\times10^{-15}$ | $(2.971\pm0.016)\times10^{-15}$ |
| $K^\pm \to \pi^0\pi^0\pi^\pm$ | $0.920\times10^{-15}$ | $(0.934\pm0.027)\times10^{-15}$ |
| $K_L^0 \to 3\pi^0$ | $2.722\times10^{-15}$ | $(2.518\pm0.089)\times10^{-15}$ |
| $K_L^0 \to \pi^+\pi^-\pi^0$ | $1.503\times10^{-15}$ | $(1.617\pm0.030)\times10^{-15}$ |
| $K^\pm \to \pi^\pm\pi^\mp e^\pm \nu$ [15] | $2.014\times10^{-18}$ | $(2.174\pm0.080)\times10^{-18}$ |
| $K^\pm \to \pi^\pm\pi^\mp \mu^\pm \nu$ | $0.672\times10^{-18}$ | $(0.744\pm0.478)\times10^{-18}$ |
| $K^\pm \to \pi^0\pi^0 e^\pm \nu$ | $1.070\times10^{-18}$ | $(1.169\pm0.213)\times10^{-18}$ |
| $K^\pm \to \pi^0\pi^0 \mu^\pm \nu$ | $0.388\times10^{-18}$ | |
| $K_L^0 \to \pi^0\pi^\pm e^\mp \nu$ | $2.126\times10^{-18}$ | $(2.764\pm0.058)\times10^{-18}$ |
| $K_L^0 \to \pi^0\pi^\pm \mu^\mp \nu$ | $0.764\times10^{-18}$ | |
| $\Lambda \to pe\nu$ [4] | $2.085\times10^{-15}$ | $(2.081\pm0.035)\times10^{-15}$ |
| $\Lambda \to p\mu\nu$ | $0.378\times10^{-15}$ | $(0.393\pm0.088)\times10^{-15}$ |
| $\Sigma^- \to ne\nu$ | $4.777\times10^{-15}$ | $(4.526\pm0.178)\times10^{-15}$ |
| $\Sigma^- \to n\mu\nu$ | $1.839\times10^{-15}$ | $(2.003\pm0.178)\times10^{-15}$ |

For the three-body decays of $A' \to Aev(\Delta S = 0)$, their width formula is:

$$\Gamma = C|\psi_{6,6}|^2 = G^2 m_i^5 (1-x^2)^6 . \qquad (28)$$

Eq.(28) includes $\pi^\pm \to \pi^0 e\nu$, $n \to pe\nu$, and $\mu \to e\bar{\nu}\nu_\mu$, $\Sigma^\pm \to \Lambda e^\pm \nu$, here $G^2 = G_0^2/192\pi^3$; $\tau \to e\bar{\nu}\nu_\tau, \to \mu\bar{\nu}_\mu\nu_\tau$, here $G^2 = G_0^2/194\pi^3$. Both G are approximately equal.

| | Calculated | Observed(MeV) | |
|---|---|---|---|
| $\pi^\pm \to \pi^0 e\nu$ | $2.671\times10^{-22}$ | $(2.619\pm0.177)\times10^{-22}$ | $G^2 = G_0^2/32\pi^3$ [2] |
| $n \to pe\nu$ | $7.2825\times10^{-25}$ =903.826sec | $(885.7\pm0.8)$sec | $G^2 = (G_0^2/\pi^3)2(1+3\alpha^2)$ and $\alpha$ =1.21 |
| $\mu \to e\bar{\nu}\nu_\mu$ | $2.9689\times10^{-16}$ | $(2.954\pm0.042)\times10^{-16}$ | $G^2 = G_0^2/192\pi^3$ [5] |
| $\Sigma^+ \to \Lambda e^+\nu$ | $1.553\times10^{-16}$ | $(1.642\pm0.410)\times10^{-16}$ | |
| $\Sigma^- \to \Lambda e^-\nu$ | $2.815\times10^{-16}$ | $(2.550\pm0.120)\times10^{-16}$ | |
| $\tau \to e\bar{\nu}\nu_\tau$ | $4.035\times10^{-10}$ | $(4.041\pm0.012)\times10^{-10}$ | $G^2 = G_0^2/194\pi^3$ |
| $\tau \to \mu\bar{\nu}_\mu\nu_\tau$ | $3.950\times10^{-10}$ | $(3.932\pm0.012)\times10^{-10}$ | |

For the three-body decays of $\eta$, their width formula is:

$$\Gamma = C|\psi_{6,6}|^2 = \frac{G^2}{4\pi} m_i (1-x^2)^6 . \qquad (29)$$

Eq.(29) includes $\eta \to 3\pi^0$ and $\eta \to \pi^+\pi^-\pi^0$, here $G = g_\eta$ =3.335; $\eta \to \pi^+\pi^-\gamma$, here

$G = g_\eta = 2.9\times10^{-3}$.

| | Calculated | Observed(MeV) |
|---|---|---|
| $\eta\to3\pi^0$ [47] | $4.188\times10^{-4}$ | $(4.226\pm0.036)\times10^{-4}$ |
| $\eta\to\pi^+\pi^-\pi^0$ | $2.975\times10^{-4}$ | $(2.951\pm0.052)\times10^{-4}$ |
| $\eta\to\pi^+\pi^-\gamma$ | $6.020\times10^{-5}$ | $(6.097\pm0.134)\times10^{-5}$ |

For various radiative decays, the known formulas are:

$$\Gamma = C\,|\psi_{3,3}|^2 = (G^2/4\pi)m_i^3(1-x^2)^3. \qquad (30)$$

Eqs.(30) include the scalar mesons $S\to\gamma\gamma$ [14], here x=0 and $G = f_0/2$; the pseudo-scalar mesons $P\to\gamma\gamma$ [3,9,14], here x=0 and *G*=*f*, e.g., $\pi^0,\eta,\eta'$; the tensor mesons $T\to\gamma\gamma$ here x=0 and $G = g/(2\sqrt{5})$ [14]; the vector mesons $V\to\varphi\gamma$ [9,19], here $G = f_1/2$; the hyperons $H\to A\gamma(\Delta S=1)$ here $G = g/\sqrt{2}$, e.g., $\Lambda\to n\gamma$, $\Sigma^+\to p\gamma$, $\Xi^0\to\Lambda\gamma,\Sigma^0\gamma$ and $\Xi^-\to\Sigma^-\gamma$, etc.[3,10,18]. Various coupling constants of these decays are different.

## 4. Possible dynamic basis

The decays always are that a particle of the initial state decays to several particles owing to the internal interaction. It points out that the unstable particles have the internal interaction, namely, have necessarily the internal structure, including $\mu^\pm$. But the initial state for the level of particle structure is always a free-particle which has not interaction with other, so the equation is the free field equation. In the momentum representation, it is:

$$\frac{\vec{r}^{\,2}}{\hbar^2}\psi + \nabla_p^2\psi = 0. \qquad (31)$$

This is namely the Helmholtz equation which takes the momentum as a variable. For leptons and baryons there should be Dirac equations, and for mesons there should be the Klein-Gorden equation. The Schrodinger equation, which has unified baryons and mesons, is also Eq.(31). In the system of spherical coordinates ($\hbar$ =1),

$$\frac{1}{p^2}[\frac{\partial}{\partial p}(p^2\frac{\partial\psi}{\partial p}) + \frac{1}{\sin\theta}\frac{\partial}{\partial\theta}(\sin\theta\frac{\partial\psi}{\partial\theta}) + \frac{1}{\sin^2\theta}\frac{\partial^2\psi}{\partial\varphi^2}] + 2\tau T\psi = 0. \qquad (32)$$

Its solution is:

$$\psi = R(p)\Theta(\theta)\Phi(\varphi). \qquad (33)$$

We suppose that the particles, which have the internal structure, are analogy with a rigid rotator, so the wave equation is:

$$\frac{1}{\sin\theta}\frac{\partial}{\partial\theta}(\sin\theta\frac{\partial\psi}{\partial\theta})+\frac{1}{\sin^2\theta}\frac{\partial^2\psi}{\partial\varphi^2}+8\pi^2 IT\psi=0. \quad (34)$$

Here $T=\frac{1}{8\pi^2 I}l(l+1)$. It is just the spherical function equation. Its solution is:

$$\psi_{l,m}=N_{lm}P_l^{|m|}(\cos\theta)e^{im\varphi}. \quad (35)$$

So $$|\psi_{l,m}|^2=\frac{1}{4\pi}\frac{(l-|m|)!(2l+1)}{(l+|m|)!}|P_l^{|m|}(\cos\theta)|^2. \quad (36)$$

The transition amplitude $S_{fi}$ may be thought to be a particular solution of the wave function, the decay rate $\Gamma$ is proportional to $|S_{fi}|^2$, and they are all probability. Then we may suppose

$$\Gamma=A|\psi_{l,m}|^2=\frac{A}{4\pi}\frac{(l-|m|)!(2l+1)}{(l+|m|)!}|P_l^{|m|}(\cos\theta)|^2. \quad (37)$$

In a system of spherical coordinates of the momentum representation, $x=\cos\theta=p_z/\vec{p}$. At the mass surface, let $\vec{p}$ is the $m_i$ of the initial state, and $p_z$ is the mass-sum of the particles of the final state $\sum m_f$. So

$$x=\frac{\sum m_f}{m_i}. \quad (38)$$

It is namely Eq.(9). Such the square of solution of the equation, i.e., some types of the associated Legendre functions $P_l^l$ and $P_{l+1}^l$, is combined with the quantum numbers conservation and the $\Delta$I=1/2 rule, we can unify many decay modes of $(1/2)^+$ lepton-baryons and $0^{-+}$ mesons, and some decays of resonances. Let $u=x^2$, so the formula of decay width

$$\Gamma=Cu^{\alpha-1}(1-u)^l,\ (\alpha=1,2) \quad (39)$$

is just the B distribution [52].

According to the selection rules, $\Delta$I=1/2 and $\Delta$S=$\Delta$Q determine N in formulas, for example, N=2 for $\Lambda\to n\pi^0$ [3]. N=2 for square of hyperon nonleptonic decay amplitudes [3]. The quantum number CP=$\pm$1 determines $K_S^0\to 2\pi, K_L^0\to 3\pi$ [3]. A radiative formula is [3]:

$$\Gamma(A\to B+\gamma)=C(\frac{M_A^2-M_B^2}{M_pM_A})^3=C\frac{M_A^3}{M_p^3}(1-x^2)^3=AM_A^3|\psi_{3,3}|^2. \quad (40)$$

In the unified decay theory, the final wave function must be the orthogonal normalization,

$$\int \psi^* \psi d\Omega = 1. \qquad (41)$$

It corresponds to that the sum of various fractions for a total decay mode should be 1. Let $|\psi^*\psi| = A\Gamma_i = R_i, \therefore \int A\Gamma_i d\Omega = 1$. $\Gamma_i$ are dispersed, so the fraction ratio $\sum_i R_i = 1$. For a particle, various decay modes should be the same equation and wave function, only l, m and k are different.

This formula may extend to a general decay formula:

$$\Gamma = A|\psi_{a,b}|^2 = Bx^m(1-x^2)^n. \qquad (42)$$

Here $x = \frac{\sum m_f}{m_i}$, $m_i$ and $\sum m_f$ are respectively mass of a particle A' of the initial state and the mass-sum of the particles of the final state. At high energy it becomes possibly a Gamma distribution.

## 5. General discussion of the decay modes and fractions

Usual decay processes classifies strong, electromagnetic and weak decays [3], in which the coupling constants are very big different. For the weak decays there are the leptonic decays, semileptonic decays and nonleptonic decays, or the two-body and three-body decays, etc. Theoretically, the general decays, except the strong decays, should be obtained from the Weinberg-Salam electroweak unification theory, and the weak decays may be obtained from the Yang-Mills gauge theory with SU(2) symmetry. Various electromagnetic decays, which seem to have $\alpha$ =1/137, should be able to be determined by QED. The strong decays of the resonances are $|\Delta$S|=0.

It is known that the quantitative level of decays is determined by the coupling constants of different interaction. For the decay modes with the same level, the fractions are determined by mass and quantum number.

The selection rule is a summation for the final particles. For example, $\Omega^- \to \Xi\pi$ and $\Lambda \to N\pi$ are $|\Delta$I|=3/2 and $|\Delta$S|=1. For $\Delta$S=0, $\Delta \to N\pi$ is $\Delta$I=0; $\Sigma \to \Lambda\gamma$ is $|\Delta$I|=1; $\eta \to 3\pi$ is $|\Delta$I|=3.

Various possible decay modes are determined from the type of interactions of decay, the conversation law of mass and charge, and the selection rules $|\Delta$S|=1 and $|\Delta$I|=1/2, etc. The lifetime and decay for different particles are classified. Usual quantitative levels of the fractions for the same type of decays are equal, for example, these lifetimes are all $10^{-10}$s for hyperons and $K_S^0, \Omega^-, A' \to A\pi$, etc. Further, various fractions are determined quantitatively from SU(3) and the quantum field theory.

We propose, firstly, the lifetimes of hyperons are determined by the lifetime formulas [21,53].

Next, the main decay modes for baryons are $H' \to H\pi$, and the fractions of $\Lambda \to p\pi^-, n\pi^0$ and $\Sigma^+ \to p\pi^0, n\pi^+$ are determined from the selection rules combined difference of masses of initial and final states. Thirdly, different classes of decay modes are researched, for example, $\Lambda \to p\pi^-, n\pi^0$ (first class), $\to n\gamma, pl^-v, p\pi^-\gamma$ (second class), $\to ne^+e^-, n\mu^+\mu^-$ (third class), etc.

For the weak decays, the selection rules are $|\Delta S|$ =0, and $|\Delta I| = |\Delta I_3|$ =1, so the $\beta$-decays are necessarily $A \to Be\nu$.

Lepton $\mu^- \to e^- v_\mu \bar{v}_e (2.19703 \pm 0.00004) \times 10^{-6} s$;

Meson $\pi^- \to \pi^0 e^- \bar{v}_e (2.6680 \pm 0.0024) \times 10^{-16} s$;

Neutron $n \to pe^- \bar{v}_e (885.7 \pm 0.8) s$.

They are similar each other, and are the $\beta$ decays of three metastable particles of the ground states in leptons, mesons and baryons, respectively.

For the decays of these smaller fractions may be:

1.According to the fractions from big to small, the radiative decays are [1]:

$\mu^+ \to e^+ \bar{v}_\mu v_e \gamma (1.4 \pm 0.4)\%$; $\tau^+ \to e^+ \bar{v}_\tau v_e \gamma (1.75 \pm 0.18)\%$.

$K^+ \to \mu^+ \bar{v}_\mu \gamma (5.50 \pm 0.28) \times 10^{-3}$; $\tau^+ \to \mu^+ \bar{v}_\tau v_\mu \gamma (3.6 \pm 0.4) \times 10^{-3}$;

$K_L^0 \to \pi^+ e^- \bar{v}_e \gamma (3.53 \pm 0.06) \times 10^{-3}$; $K_S^0 \to \pi^+ \pi^- \gamma (1.78 \pm 0.05) \times 10^{-3}$;

$\Xi^0 \to \Sigma^0 \gamma (3.33 \pm 0.10) \times 10^{-3}; \to \Lambda\gamma (1.18 \pm 0.30) \times 10^{-3}$;

$\Sigma^+ \to p\gamma (1.23 \pm 0.05) \times 10^{-3}$.

$\Lambda \to p\pi^- \gamma (8.4 \pm 1.4) \times 10^{-4}$; $\Sigma^- \to n\pi^- \gamma (4.6 \pm 0.6) \times 10^{-4}$;

$\Sigma^+ \to n\pi^+ \gamma (4.5 \pm 0.5) \times 10^{-4}$; $K_L^0 \to \gamma\gamma (5.96 \pm 0.15) \times 10^{-4}$;

$\pi^+ \to \mu^+ \bar{v}_\mu \gamma (2.00 \pm 0.25) \times 10^{-4}$;

$K^+ \to \pi^0 \pi^+ \gamma (2.75 \pm 0.15) \times 10^{-4}; \to \pi^0 e^+ v_e \gamma (2.65 \pm 0.20) \times 10^{-4}$;
$\to 2\pi^+ \pi^- \gamma (1.40 \pm 0.31) \times 10^{-4}$;

$\Xi^- \to \Sigma^- \gamma (1.27 \pm 0.23) \times 10^{-4}$. $K_L^0 \to \pi^+ \pi^- \gamma (4.38 \pm 0.13) \times 10^{-5}$.

$\pi^+ \to e^+ v_e \gamma (1.61 \pm 0.23) \times 10^{-7}$.

2.According to the fractions from big to small, the $\beta$ decays are [1]:

$\Omega^- \to \Xi^0 e^- \overline{v}_e (5.6 \pm 2.8) \times 10^{-3}$; $\Xi^0 \to \Sigma^+ e^- \overline{v}_e (2.7 \pm 0.4) \times 10^{-3}$.

$\Sigma^- \to n e^- \overline{v}_e (10.17 \pm 0.34) \times 10^{-4}; \to n \mu^- \overline{v}_\mu (4.5 \pm 0.4) \times 10^{-4}$;

$\Lambda \to p e^- \overline{v}_e (8.32 \pm 0.14) \times 10^{-4}; \to p \mu^- \overline{v}_\mu (1.57 \pm 0.35) \times 10^{-4}$;

$\Xi^- \to \Lambda e^- \overline{v}_e (5.63 \pm 0.30) \times 10^{-4}; \to \Lambda \mu^- \overline{v}_\mu (3.5 \pm 2.8) \times 10^{-4}$;

$\pi^+ \to e^+ v_e (1.230 \pm 0.004) \times 10^{-4}$.

$\Sigma^- \to \Lambda e^- \overline{v}_e (5.73 \pm 0.27) \times 10^{-5}$; $\Sigma^+ \to \Lambda e^+ v_e (2.0 \pm 0.5) \times 10^{-5}$;

$K^+ \to \pi^+ \pi^- e^+ v_e (4.08 \pm 0.09) \times 10^{-5}; \to \pi^+ \pi^- \mu^+ v_\mu (1.4 \pm 0.9) \times 10^{-5}$;

$K^+ \to e^+ v_e (1.55 \pm 0.07) \times 10^{-5}; \to 2\pi^0 e^+ v_e (2.1 \pm 0.4) \times 10^{-5}$.

$\pi^+ \to \pi^0 e^+ v_e (1.025 \pm 0.034) \times 10^{-8}$.

3.According to the fractions from big to small, the decays of pair-lepton are [1]:

$\Sigma^0 \to \Lambda e^+ e^- (5 \times 10^{-3})$. $\pi^0 \to e^+ e^- e^+ e^- (3.14 \pm 0.30) \times 10^{-5}$.

$K^+ \to \pi^+ e^- e^+ (2.88 \pm 0.13) \times 10^{-7}; \to e^+ e^- \mu^+ v_\mu (1.3 \pm 0.4) \times 10^{-7}$.

$K_L^0 \to \mu^+ \mu^- (7.25 \pm 0.16) \times 10^{-9}$.

For the leptonic decays $\pi^\pm, K^\pm \to lv$, the formulas are symmetric, and their other similar decays are possibly also symmetric.

We think that the decays may be determined from the lifetime formulas [21,53], the selection rules and the unified decay formulas. For fermions, the weak decays of hyperons are all the two-body decays. The decays of leptons and neutron, and the leptonic decays of hyperons may be calculated. Other radiative decays of hyperons are all smaller fractions. The main decays of bosons include $\pi$ mesons, $K^\pm$ and $K^0$ mesons, and other mesons. The decays of all hadrons, firstly, pass nearly through the strong decay to the same kind of the ground states, and then pass through radiative and weak decays to nucleons and n→p, or to $\pi \to \mu \to e$.

Various decay modes of resonances are more definite. Except the radiative decays of some mesons, they are mainly the strong decay, so various quantum numbers are invariance, and obey the SU(3). For baryon resonances, $\Delta(1232) \to N\pi$, $\Lambda(1405) \to \Sigma\pi$ and $\Xi(1530) \to \Xi\pi$, etc., have 100% fractions. For meson resonances, $\rho(770) \to \pi\pi$ and $K^*(892) \to K\pi$, etc., have approximately 100%.

We proposed the width formula [21]:

$$\Gamma = aJ + b, \qquad (43)$$

which corresponds to a mass formula of Regge theory:

$$S=AJ+B. \qquad (44)$$

We suggested also that the mass formulas of resonances for mesons and baryons with the same $J^{PC}$ are equal spacing [54]:

$$M = M_0 + BI, \quad M = M_0 + AS. \qquad (45)$$

The decays of resonances seem to be different classes. For instance, before N*→ $N\pi$, after N*→ $N\pi\pi$ and so on. The decay modes $e^+e^- / \mu^+\mu^-$ of resonances are approximately equal, for example, $\Gamma(v^0 \to l^+l^-)$ [1,3].

If a coupling constant determines only a decay, for instance, $\Sigma^0 \to \Lambda + \gamma$, the meaning of formulas will be very finite. They should extend to various similar radiative decays. For example,

$\rho \to \dfrac{\pi^0\gamma(6.0 \pm 0.8)}{\pi^{\pm}\gamma(4.5 \pm 0.5)}$=4/3, $\rho \to \pi\gamma(149.4 \pm 1.0) \times (10.5 \pm 1.3)$ =156.87KeV.

$\omega \to \pi^0\gamma(84.9 \pm 0.8) \times (8.9 \pm 0.2)$ =755.61KeV.

$\Gamma(\rho,\omega,\phi \to e^+e^-)$=7.0218：0.6096：0.3110KeV=23：2：1.

The widths of $\pi^0,\eta \to \gamma\gamma$ and $\eta' \to \gamma\gamma$ are respectively 7.742, 520 and 4300eV. Here x=0, and $\Gamma(\pi^0 \to \gamma\gamma) = Am_\pi^3$ [3]. It is extended to $\pi^0,\eta \to \gamma\gamma$, and both constants are the same. But, it is not agree with $\eta' \to \gamma\gamma$.

Some fractions for the particles of many decay modes are ratio of equality [22]. For example,

$\Sigma^- \to n\pi^-, \to nev : n\mu v[(10.17 \pm 0.34) : (4.5 \pm 0.4)10^{-4}]$=2;

$\Xi^- \to \Lambda\pi^-, \to \Lambda ev : \Lambda\mu v[(5.63 \pm 0.31) : (3.5 \pm 3)10^{-4}]$.

$\Lambda \to plv / p\pi^- = [(0.989 \pm 0.157) / 0.639] \times 10^{-3}$;

$\Sigma^- \to nlv / n\pi^- = (1.47 \pm 0.07) \times 10^{-3}$;

$\Xi^- \to \Lambda lv / \Lambda\pi^- = (0.913 \pm 0.6) \times 10^{-3}$; these values are approximately $1.5 \times 10^{-3}$.

## 6. Search for decays of massive hadrons

The fractions of $\Omega^- \to \Lambda K^- : \Xi^0\pi^- : \Xi^-\pi^0$ are approximately 3:1 and 3:1. They may be summarized to the formulas:

| | $\Gamma$ | Calculated | Observed(MeV) |
|---|---|---|---|
| $\Omega^- \to \Lambda K^-$ | $\frac{G^2}{4\pi} m_\Omega^2 (1-x^2)^3$ | $5.418\times10^{-12}$ | $(5.436\pm0.056)\times10^{-12}$ |
| $\Omega^- \to \Xi^0\pi^-$ | $\frac{G^2}{4\pi N} m_\Omega (1-x^2)$ (N=1) | $1.947\times10^{-12}$ | $(1.892\pm0.056)\times10^{-12}$ |
| $\Omega^- \to \Xi^-\pi^0$ | $\frac{G^2}{4\pi N} m_\Omega (1-x^2)$ (N=3) | $6.436\times10^{-13}$ | $(6.895\pm0.321)\times10^{-13}$ |

For hadrons of C=1, $\Lambda_c^+ \to p\overline{K}^0$ , $\Lambda_c^+ \to pK^-\pi^+$ , $\Lambda_c^+ \to p\overline{K}^0\pi^0$ , $\Lambda_c^+ \to p\overline{K}^0\eta$ ( $G^2 = 0.25$ ) and $D^0 \to K^- e^+ \nu_e$ , $D^0 \to K^-\mu^+\nu_\mu$ , $D^0 \to \overline{K}*(892)^- e^+\nu_e$ , $D^0 \to \overline{K}*(892)^- \mu^+\nu_\mu$ ( $G^2 = 0.042$ ) obey the formula (27), which agree within the errors.

| | Calculated | Observed(MeV) |
|---|---|---|
| $\Lambda_c^+ \to p\overline{K}^0$ | $8.158\times10^{-11}$ (N=3) | $(7.589\pm1.975)\times10^{-11}$ |
| $\Lambda_c^+ \to pK^-\pi^+$ | $12.66\times10^{-11}$ | $(16.455\pm4.278)\times10^{-11}$ |
| $\Lambda_c^+ \to p\overline{K}^0\pi^0$ | $12.71\times10^{-11}$ | $(10.860\pm3.291)\times10^{-11}$ |
| $\Lambda_c^+ \to p\overline{K}^0\eta$ | $2.786\times10^{-11}$ | $(3.949\pm1.316)\times10^{-11}$ |
| $D^0 \to K^- e^+\nu_e$ | $5.386\times10^{-11}$ | $(5.634\pm0.177)\times10^{-11}$ |
| $D^0 \to K^-\mu^+\nu_\mu$ | $5.011\times10^{-11}$ | $(5.120\pm0.257)\times10^{-11}$ |
| $D^0 \to \overline{K}*(892)^- e^+\nu_e$ | $3.705\times10^{-11}$ | $(3.483\pm0.257)\times10^{-11}$ |
| $D^0 \to \overline{K}*(892)^- \mu^+\nu_\mu$ | $3.176\times10^{-11}$ | $(3.130\pm0.401)\times10^{-11}$ |

$D^+ \to \overline{K}^0 e^+\nu_e$ , $D^+ \to \overline{K}^0\mu^+\nu_\mu$ , $D^+ \to \overline{K}*(892)^0 e^+\nu_e$ , $D^+ \to \overline{K}*(892)^0\mu^+\nu_\mu$ ( $G^2 = 0.042$ ) obey the formula (25), which agree within the errors.

| | Calculated | Observed(MeV) |
|---|---|---|
| $D^+ \to \overline{K}^0 e^+ \nu_e$ | $5.804\times10^{-11}$ | $(5.443\pm0.316)\times10^{-11}$ |
| $D^+ \to \overline{K}^0 \mu^+ \nu_\mu$ | $5.597\times10^{-11}$ | $(6.013\pm0.506)\times10^{-11}$ |
| $D^+ \to \overline{K}*(892)^0 e^+ \nu_e$ | $2.405\times10^{-11}$ (N=2) | $(2.367\pm0.133)\times10^{-11}$ |
| $D^+ \to \overline{K}*(892)^0 \mu^+ \nu_\mu$ | $2.227\times10^{-11}$ (N=2) | $(2.342\pm0.190)\times10^{-11}$ |

The fractions of $W^+ \to e^+\nu, \mu^+\nu, \tau^+\nu$ are about equal. It shows the leptonic universality.

In this article we do not discuss the small decay modes, whose fractions are not definite at present, and the decay rates of the general resonances.

**References**

1.W.-M Yao, C.Amsler, D.Asner, et al. Particle Data Group. J.Phys. G33,1(2006).
2.R.E.Marshak, Riazuddin and C.P.Ryan, Theory of Weak Interactions in Particle Physics. Wiley-Interscience. 1969.
3.B.T.Feld, Models of Elementary Particle. Blaisdell Publishing Company. 1969.
4.N.Cabibbo, Phys.Rev.Lett. 10,531(1963).
5.R.P.Feynman and M.Gell-Mann, Phys.Rev. 109,193(1958).
6.L.M.Chounet, J.-M.Gaillard and M.-K.Gaillard, Phys.Reports. 4,199(1972).
7.J.Smith and Z.E.S.Uy, Phys.Rev. D8,3056(1973).
8.B.R.Holstein, Rev.Mod.Phys. 46,789(1974).
9.A.Bramon and M.Greco, Phys.Lett. 48B,137(1974).
10.J.K.Bajaj and A.K.Kapoor, Phys.Rev. D16,2226(1977).
11.Y.Abe, K.Fujii and K.Sato, Prog.Theor.Phys. 58,1849(1977).
12.D.Bryman and C.Picciotto, Rev.Mod.Phys. 50,11(1978).
13.A.Sirlin, Rev.Mod.Phys. 50,573(1978).
14.M.Greco and Y.Srivastava, Nuov.Cim. 43A,88(1978).
15.K.Ozaki, Prog.Theor.Phys. 59,515(1978).
16.T.J.Devlin and J.O.Dickey, Rev.Mod.Phys. 51,237(1979).
17.P.J.O'Donnell, Rev.Mod.Phys. 53,673(1981).
18.S.Minami and R.Nakashima, Lett.Nuov.Cim. 30,293(1981).
19.S.Minami and R.Nakashima, Prog.Theor.Phys. 65,254(1981).
20.E.Kaxiras, E.J.Moniz and M.Soyeur, Phys.Rev. D32,695(1985).
21.Yi-Fang Chang, New Research of Particle Physics and Relativity. Yunnan Science and Technology Press. 1989. Phys.Abst. 93,No.1371(1990).
22.Yi-Fang Chang, J.Xinjiang University. 7,3,43(1990).
23.J.L.Ritchie and S.G.Wojcicki, Rev.Mod.Phys. 65,1149(1993).
24.J.D.Richman and P.R.Burchat, Rev.Mod.Phys. 67,893(1995).
25.L.R.Surguladze and M.A.Samuel, Rev.Mod.Phys. 68,259(1996).
26.G.Buchalla, A.J.Buras and M.E.Lautenbacher, Rev.Mod.Phys. 68,1125(1996).
27.Y.Kuno and Y.Okada, Rev.Mod.Phys. 73,151(2001).
28.T.Hurth, Rev.Mod.Phys. 75,1159(2003).

29.N.Severijns, M.Beck and O.Naviliat-Cuncic, Rev.Mod.Phys. 78,991(2006).
30.M.Davier, A.Hocker and Z.Zhang, Rev.Mod.Phys. 78,1043(2006).
31.T.Ibrahim and P.Nath, Rev.Mod.Phys. 80,577(2008).
32.J.Insler, et al. (CLEO Collaboration), Phys.Rev. D81,091101(2010).
33.M.Petric, et al. (The Belle Collaboration), Phys.Rev. D81,091102(2010).
34.G.Bozzi, F.Campanario, V.Hankele and D.Zeppenfeld, Phys.Rev. D81,094030(2010).
35.O.Leitner, J.-P.Dedonder, B.Loiseau and R.Kaminski, Phys.Rev. D81,094033(2010).
36.S.Davidson and G.Grenier, Phys.Rev. D81,095016(2010).
37.J.P.Lees, et al. (BABAR Collaboration), Phys.Rev. D81,111101(2010).
38.R.A.Briere, et al., Phys.Rev. D81, 112001(2010).
39.W.Rodejohann, Phys.Rev. D81,114001(2010).
40.H.-Y.Cheng and C.-K.Chua, Phys.Rev. D81,114006(2010).
41.S.Dubnicka, A.Z.Dubnickova, M.A.Ivanov and J.G.Korner, Phys.Rev. D81,114007(2010).
42.T.D.Lee, Particle Physics and Introduction to Field Theory. Harwood Academic Publishers. 1981.
43.E.M.Lifshitz and L.P.Pitaevskii, Relativistic Quantum Theory (Part 2). Pergamon Press. 1974.
44.W.Greiner and B.Muller, Gauge Theory of Weak Interactions(3 Edition). Springer-Verlag. 2002.
45.S.Weinberg, The Quantum Theory of Fields. V2. Cambridge. 2001.
46.R.J.Oakes, Phys.Lett. 30B,262(1969).
47.A.Nandy, V.P.Gautom and B.Bagchi, Phys.Lett. 82B,121(1979).
48.R.Jacob and R.G.Sachs, Phys.Rev. 121,350(1961).
49.M.Bander, D.Silverman and A.Soni, Phys.Rev.Lett. 44,7(1980).
50.L.M.Sehgal and J.van Leusen, Phys.Lett. B483,300(2000).
51.N.Cabibbo, E.C.Swallow and R.Winston, Phys.Rev.Lett. 25,251803(2004).
52.Yi-Fang Chang, Hadronic J. 7,1118(1984).
53.Yi-Fang Chang, Y-Q and I-U Symmetries, and New Unified Lifetime Formulas of Hadrons. arXiv:0907.2763.
54.Yi-Fang Chang, J.Yunnan University. 11,2,125(1989).